\documentclass{article}

\usepackage{arxiv}

\usepackage[utf8]{inputenc} 
\usepackage[T1]{fontenc}    
\usepackage{hyperref}       
\usepackage{url}            
\usepackage{booktabs}       
\usepackage{amsfonts}       
\usepackage{nicefrac}       
\usepackage{microtype}      
\usepackage{lipsum}
\usepackage{graphicx}
\graphicspath{ {./images/} }

\title{Transmission Structured Illumination Microscopy using Tilt-mirror Assembly}

\author{\href{https://orcid.org/0000-0002-9657-6519}{\includegraphics[scale=0.06]{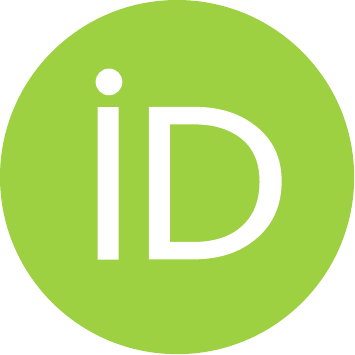}\hspace{1mm}
 Krishnendu Samanta} \\
  Department of Physics\\
  Indian Institute of Technology Delhi\\
  New Delhi 110016, India \\
   \And
   \href{https://orcid.org/0000-0001-5372-1571}{\includegraphics[scale=0.06]{orcid.pdf}\hspace{1mm}
 Azeem Ahmad} \\
  Department of Physics and Technology\\
  UiT-The Arctic University of Norway\\
  Troms{\o} 9037, Norway \\
  \And
 Jean-Claude Tinguely \\
  Department of Physics and Technology\\
  UiT-The Arctic University of Norway\\
  Troms{\o} 9037, Norway \\
  \And
  \href{https://orcid.org/0000-0001-7841-6952}{\includegraphics[scale=0.06]{orcid.pdf}\hspace{1mm}
 Balpreet Singh Ahluwalia} \\
  Department of Physics and Technology\\
  UiT-The Arctic University of Norway\\
  Troms{\o} 9037, Norway \\
  \texttt{balpreet.singh.ahluwalia@uit.no} \\
  \And
  \href{https://orcid.org/0000-0002-1513-8008}{\includegraphics[scale=0.06]{orcid.pdf}\hspace{1mm}
 Joby Joseph} \\
  Department of Physics\\
  Indian Institute of Technology Delhi\\
  New Delhi 110016, India \\
  \texttt{joby@physics.iitd.ac.in} \\
}

\begin{document}
\maketitle
\begin{abstract}
We present experimental demonstration of tilt-mirror assisted transmission structured illumination microscopy (tSIM) that offers a large field of view super resolution imaging. An assembly of custom-designed tilt-mirrors are employed as the illumination module where the sample is excited with the interference of two beams reflected from the opposite pair of mirror facets. Tunable frequency structured patterns are generated by changing the mirror-tilt angle and the hexagonal-symmetric arrangement is considered for the isotropic resolution in three orientations. Utilizing high numerical aperture (NA) objective in standard SIM provides super-resolution compromising with the field-of-view (FOV). Employing low NA (20X/0.4) objective lens detection, we experimentally demonstrate $\sim$ (0.56mm$\times$0.35mm) size single FOV image with $\sim$1.7- and $\sim$2.4-fold resolution improvement (exploiting various illumination by tuning tilt-mirrors) over the diffraction limit. The results are verified both for the fluorescent beads as well as biological samples.The tSIM geometry decouples the illumination and the collection light paths consequently enabling free change of the imaging objective lens without influencing the spatial frequency of the illumination pattern that are defined by the tilt-mirrors. The large and scalable FoV supported by tSIM  will find usage for applications where scanning large areas are necessary as in pathology and applications where images must be correlated both in space and time.
\end{abstract}


\section*{Introduction}
Breaking the diffraction limit \cite{abbe1873beitrage} of classical fluorescence microscopy in the last two decades revolutionized the biomedical studies and leading to a new field of research called `optical nanoscopy' \cite{hell2007far}. In the rapidly progressing area of nanoscopy, structured illumination microscopy (SIM) \cite{gustafsson2000surpassing,heintzmann2017super} appears as a significant widefield super-resolution technique where a series of structured patterns are employed to excite the fluorescent sample and corresponding raw Moir\'e frames are computationally processed to achieve around two-fold resolution enhancement over the widefield limit. In spite of offering relatively moderate resolution enhancement in comparison to the other optical super-resolution techniques, such as stimulated emission depletion (STED) \cite{hell1994breaking}, stochastic optical reconstruction microscopy (STORM) \cite{rust2006sub}, photo activated localization microscopy (PALM) \cite{betzig2006imaging}, or ground state depletion (GSD) \cite{folling2008fluorescence}, SIM has drawn considerable attention due to its high spatio-temporal resolution, low photo-toxicity, compatibility with common fluorescent labelling, efficient multi-colour imaging \cite{markwirth2019video,schermelleh2008subdiffraction} and so on. It is also considered a promising approach for the investigation of the sub-cellular dynamics of living cells \cite{kner2009super,li2015extended} for its requirement of less number of raw frames and low photon-dose. Although sinusoidal structured excitation pattern is the key feature to achieve the superior resolution imaging through SIM, it is no longer restricted to the periodic illumination patterns. Speckle-like \cite{mudry2012structured,bender2021circumventing} random illuminations with blind reconstruction approaches are also successfully implemented for the SIM imaging. However, the random illumination methods offer super-resolution at the cost of low temporal resolution due to the requirement of a large number of frames ($\sim$100s). These methods wipe out the main theme of the periodic structured illumination imaging that minimizes the required number of frames 9(15) in 2D(3D) SIM cases. Hence, the efforts to further improve the resolution of SIM by maintaining the well-defined periodic illumination patterns are worthwhile.

In the conventional linear SIM technique, a single objective lens is employed for illuminating the sample as well as for collecting the fluorescence signal. Consequently, both the illumination and the detection optics are diffraction limited by the same objective lens, and the system is strictly restricted to offer $\leq2$-fold resolution improvement compared to the classical diffraction limit. In order to surpass the typical SIM resolution limit further, there is total internal reflection fluorescent (TIRF)-SIM \cite{fiolka2008structured,brunstein2013full} where the high frequency interference pattern of evanescent waves illuminates the sample. However, the TIRF illumination is restricted to a thin optical section (<100 nm) and therefore deals with 2D imaging only. Besides, the saturation properties of the fluorescent materials are harnessed in the nonlinear-SIM \cite{gustafsson2005nonlinear,rego2012nonlinear,samanta2022saturable} approach which incorporates the contributions of multiple harmonics in super-resolution imaging. However, the requirement of high intensity to reach the operating saturation level of the fluorophores can provide photo-toxicity issues for the non-linear approach. Further SIM techniques based on different principles have also been reported, e.g., plasmonic \cite{wei2010plasmonic,wei2014wide,bezryadina2018high}, proximity projection grating \cite{hu2015sub} or photonic chip \cite{helle2020structured}. Each of these techniques requires sophisticated and dedicated photonic or plasmonic materials to manipulate the structured illumination pattern in the sample plane. In addition, specially designed tools are required for the phase-shifting and changing pattern orientation, e.g., thermo-optics for photonic chip SIM or galvo-scanning for the plasmonic SIM. In general, these techniques are complex due to their dependence on the dedicated material; which restricts the tunability of the illumination pattern frequency. In plasmonic SIM, the pre-calibrated array structures are fabricated according to the operating wavelength range. The specific structure selectively determines the orientation and the frequency of the illumination pattern, there is no tunability of the pattern. Similarly, in photonic chip SIM, the illumination pattern frequency is pre-decided by the angles of the available waveguide arms. Finally, conventional samples on microscope slides or coverslips do not work for these methods. It is required to prepare the sample on the surface of the designed material which is illuminated by the standing waves interference pattern in the evanescent regime, restricting these techniques to 2D-TIRF super-resolution.

In the race of achieving the best possible resolution, high numerical aperture objective lenses are typically utilized for the SIM imaging. The super-resolution image in SIM employing the high NA objective lens is obtained by compromising with the field of view (FOV). This limitation appears because of the inherent trade-off between the resolution and the FOV of the standard imaging system. Because of this dependence, the next level super-resolution techniques have been dealing with large FOV imaging while improving the resolution \cite{diekmann2017chip,engdahl2021large,joseph2019improving}. Recently, a theoretical approach \cite{joseph2019improving} was reported to overcome this limitation by decoupling the illumination path from the collection path using a transmission type tilt-mirror configuration, albeit with no experimental demonstration. Herein, we demonstrate the experimental proof-of-concept of a tilt-mirror assisted transmission type SIM technique which possesses potential to exploit the high throughput SIM utilizing the free space optics in a cost-effective way. Liberating the illumination optics from the diffraction limit of the collection objective lens facilitates the full flexibility to play with the illumination configuration. This is, to the best of our knowledge, also the first experimental proof that describes it being possible to design the illumination for large FOV nanoscopy without guiding the light through any condenser lens \cite{fiolka2008structured,forster2014simple}, waveguide chip \cite{helle2020structured,engdahl2021large} or advanced materials \cite{wei2014wide,lee2021metamaterial,ma2018experimental}. 

\section*{Results}
Decoupling the illumination from the detection optics in transillumination microscopy architecture paves the way to explore the versatile illumination modalities. In this work, we present a proof-of-principle experimental demonstration of transillumination SIM technique where an assembly of mini tilt-mirrors work as the illumination module. As the excitation path becomes independent of the collection path, the maximum frequency of the illumination pattern is no longer restricted by the detection objective lens.
\begin{figure}[t!]
\centering
{\includegraphics[width=\linewidth]{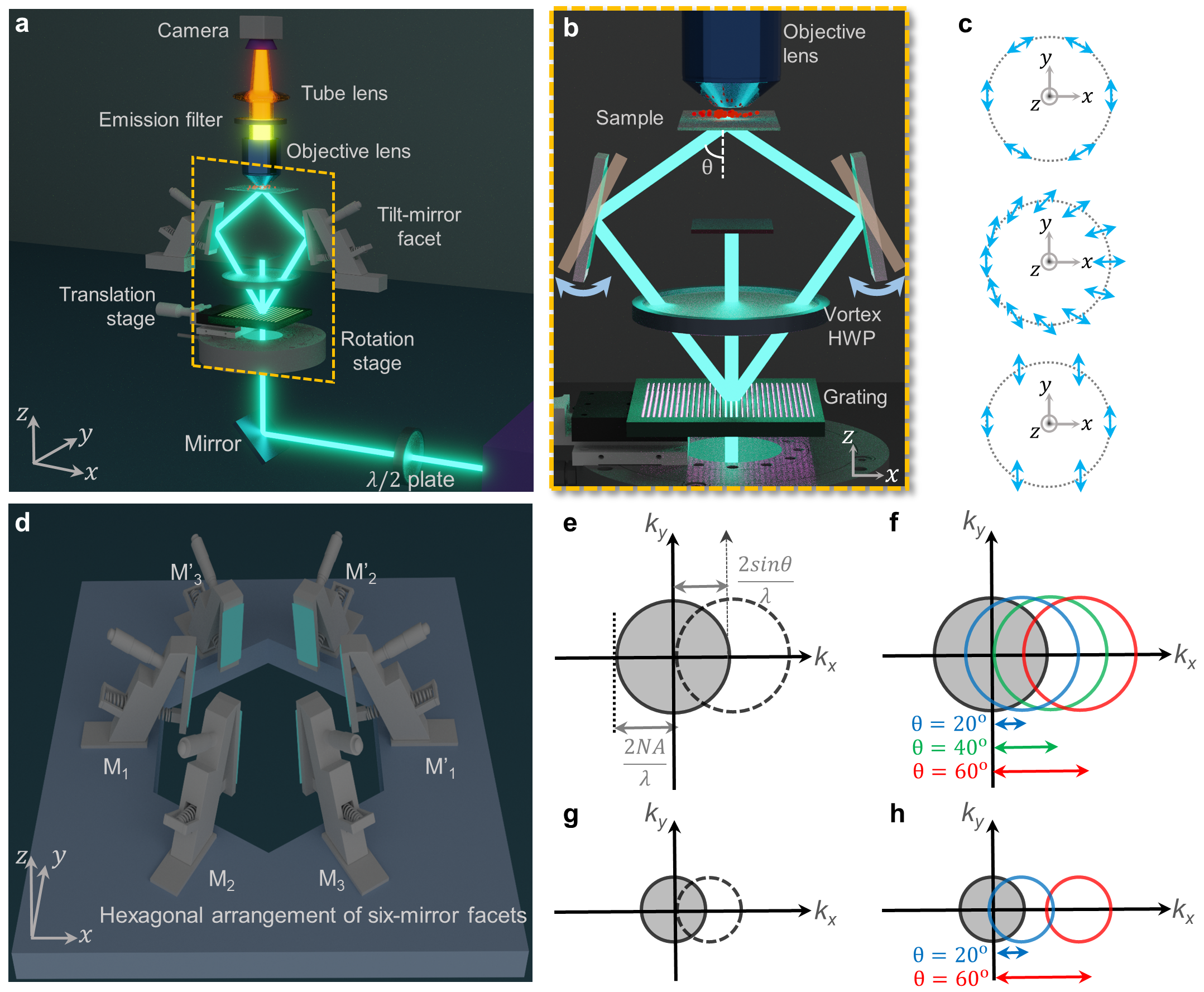}}
\caption{Concept of tSIM: (a) scheme of the experimental setup for one orientation; (b) tunable frequency illumination patterns are generated by a pair of mirror facets corresponding to low and high frequency illumination respectively; (c) scheme of polarization conversion of the interfering beams where top, middle and below panel corresponds to the state of polarization above, on and below the vortex half wave plate respectively; (d) hexagonal arrangement of mirror-facets for three orientation angles required for isotropic resolution enhancement; (e) resolution limit represented by the optical transfer function (solid circle), maximal shift (dotted circle) in conventional SIM offers 2-fold resolution enhancement. (f) maximum spectral support of tSIM governed by with periodicity $\frac{\lambda }{{2\sin {\theta_1}}}$, tunable under different interfering angles $\theta$; (g,h) frequency support of a low-NA detection objective lens under similar illumination configurations, where tSIM support more than conventional 2X resolution enhancement of SIM .}
\label{fig:setup}
\end{figure}
The graphical representation of the experimental setup is shown in Figure \ref{fig:setup}(a). To make the system compact, a physical grating is employed to split the input light, with the first order diffracted beams directed towards the optical axis ($z$) upon reflection by a pair of tilt-mirrors. The tilt angles of the oppositely faced mirror pair are adjusted in such a way that the reflected beams overlap in a small volumetric region to generate a sinusoidal interference pattern. The purpose of the condenser lens is effectively fulfilled by the set of tilt-mirrors. This unique lensless illumination scheme possesses potential benefits in generating aberration free structured pattern which is homogeneous over the entire volume of interference. Besides, the area of illumination is no longer limited by the condenser lens; rather it is decided by the diameter of the interfering beams (Supplementary 1). The sample is kept within the volumetric region to illuminate with the excitation pattern and a standard upright microscope is employed to collect the fluorescent signal from the other side of the sample. The maximum spatial frequency support of the transillumination SIM technique is determined by considering the resultant contributions from the detection and the illumination configuration,
\begin{equation}
\mathbf{k} = {\mathbf{k}_{det}} + {\mathbf{k}_{illu}}
\end{equation}
The detection system consisting of a lens with numerical aperture (NA) and emission wavelength $\lambda_{em}$ is restricted by the classical diffraction limit $\left|\mathbf{k}_{det}\right| = \frac{2\mbox{\scriptsize NA}}{\lambda_{em}}$. On the other hand, the illumination frequency $\left|\mathbf{k}_{illu}\right| = \frac{2nsin\theta}{\lambda_{ex}}$ solely depends upon the interference conditions where $\lambda_{ex}$ is the peak excitation and $\theta$ is the semi interference angle between the interfering beams within a medium of refractive index $n$. Hence, the theoretical lateral resolution limit of the tilt-mirror based tSIM is determined as,
\begin{equation}
    \Delta = \frac{\lambda_{em}}{2\left(\mbox{NA}+\frac{\lambda_{em}}{\lambda_{ex}}n\sin{\theta}\right)}
\end{equation}
The semi-angle of interference ($\theta = 2\delta + \alpha$) is governed by the diffraction angle ($\alpha$) of the grating and the inclination ($\delta$) of the mirror with respect to the optics axis. Tuning the tilt angle of the mirror leads to the change in the interference angle and also in the pattern frequency. Hence, the technique is capable to generate the tunable frequency sinusoidal structured patterns depending upon the requirement of the illumination scheme. This particular characteristic is quite unique relative to the standard SIM technologies existing so far. The close-up view within the yellow-box in Figure \ref{fig:setup}(a) is shown by Figure \ref{fig:setup}(b), representing the angle-tuning scheme for the illumination (Supplementary 1). The desired phase-shifting of the interference pattern in the sample plane is incorporated by shifting the physical grating along the lateral (along the grating period) direction with a single-axis piezo-stage. In order to create the illumination pattern of highest modulation depth, the polarization states of the interfering beams pairs are suitably converted before hitting the mirrors. Figure \ref{fig:setup}(c) represents the conversion scheme of the state of polarization: linearly polarized input (bottom panel), vortex half wave plate as polarization converter (middle panel) and azimuthally polarized output (top panel). Besides aligning the interfering beam pairs polarized parallel to each other, such conversion scheme efficiently utilizes the input power by minimizing the Fresnel reflection loss from the mirror surfaces. The description of the setup so far revolved around the excitation pattern along one orientation. The hexagonal arrangement of the tilt-mirrors are designed to create sinusoidal illumination along three different orientations for the isotropic resolution improvement. The hexa-assembly of the mirror facets are shown in the Figure \ref{fig:setup}(d), where (M$_{1}$, M$_{1}^{'}$), (M$_{2}$, M$_{2}^{'}$), (M$_{3}$, M$_{3}^{'}$) correspond to the three set of mirror-pairs. To change the illumination frequency, the equal tuning in the tilt-angles of each mirror-pair is done in such a way that the common volumetric interfering region shifts only along axial direction keeping the same lateral position. When the interference between the beam-pair occurs at a small angle ($\theta$), a large pattern period is created to provide low resolution gain whereas interference at high angle offers relatively better resolution. The resolution enhancement capability of the conventional SIM is presented by the gray circles in the Figure \ref{fig:setup}(e) that represents around two-fold gain. On the other hand, the resolution improvement of the proposed tSIM is shown by the extension of blue, green and red coloured circles in the Figure \ref{fig:setup}(f) corresponding to the semi-angle of interference $\theta = 20^{\circ}$, $40^{\circ}$ and $60^{\circ}$, respectively. The nitty-gritty details of the experimental setup is described in the Supplementary 2.

The major advantage of the tSIM technique comes into play when the large field of view imaging is concerned, where a low-magnification/low-NA objective lens is utilized for the detection. Without modifying the detection passband (keeping the objective lens fixed), it is possible to incorporate illumination pattern of any arbitrary frequency with a maximum limit of $\frac{2}{\lambda}$ in the free space optics. Figure \ref{fig:setup}(g) represents the resolution enhancement of a low-NA objective lens in traditional SIM case whereas Figure \ref{fig:setup}(h) describes the tSIM case for the illumination patterns of variable spatial frequencies. Although an illumination pattern of very high frequency (compared to the low-NA passband limit) can be achieved to provide resolution far better than conventional SIM, intermediate frequencies from lower interference angles are required to prevent a gap. By tailoring the illumination patterns with the tilt-mirror assisted scheme, the technique is capable to fill the required frequency space and to offer resolution enhancement similar to a high-NA objective lens along with a large FOV supported by a low-NA detection objective lens. 

\subsection*{tSIM imaging with conventional illumination}
In order to validate the super-resolution capability of the proposed tilt-mirror assisted structured illumination microscopy, we present experiments for polymere beads as well as a biological cell sample. For the proof-of-concept, first we demonstrate the experimental results for a scenario similar to the conventional SIM case where the illumination frequency lies within the passband limit of the detection system. A CW laser of wavelength 532 nm (DPSS 05-01, Cobolt Samba) is used as the excitation source throughout the study. The interfering beams are overlapped at the semi-angle of interference $\theta = 19^{\circ}$ to generate the sinusoidal illumination pattern of periodicity $\sim$778 nm. The interfering beams are of $\sim$1 mm in diameter to create the homogeneous illumination pattern over the entire area of illumination in the sample plane, which is sufficiently large relative to the FOV of the detection system. 
\begin{figure}[t!]
\centering
{\includegraphics[width=1\linewidth]{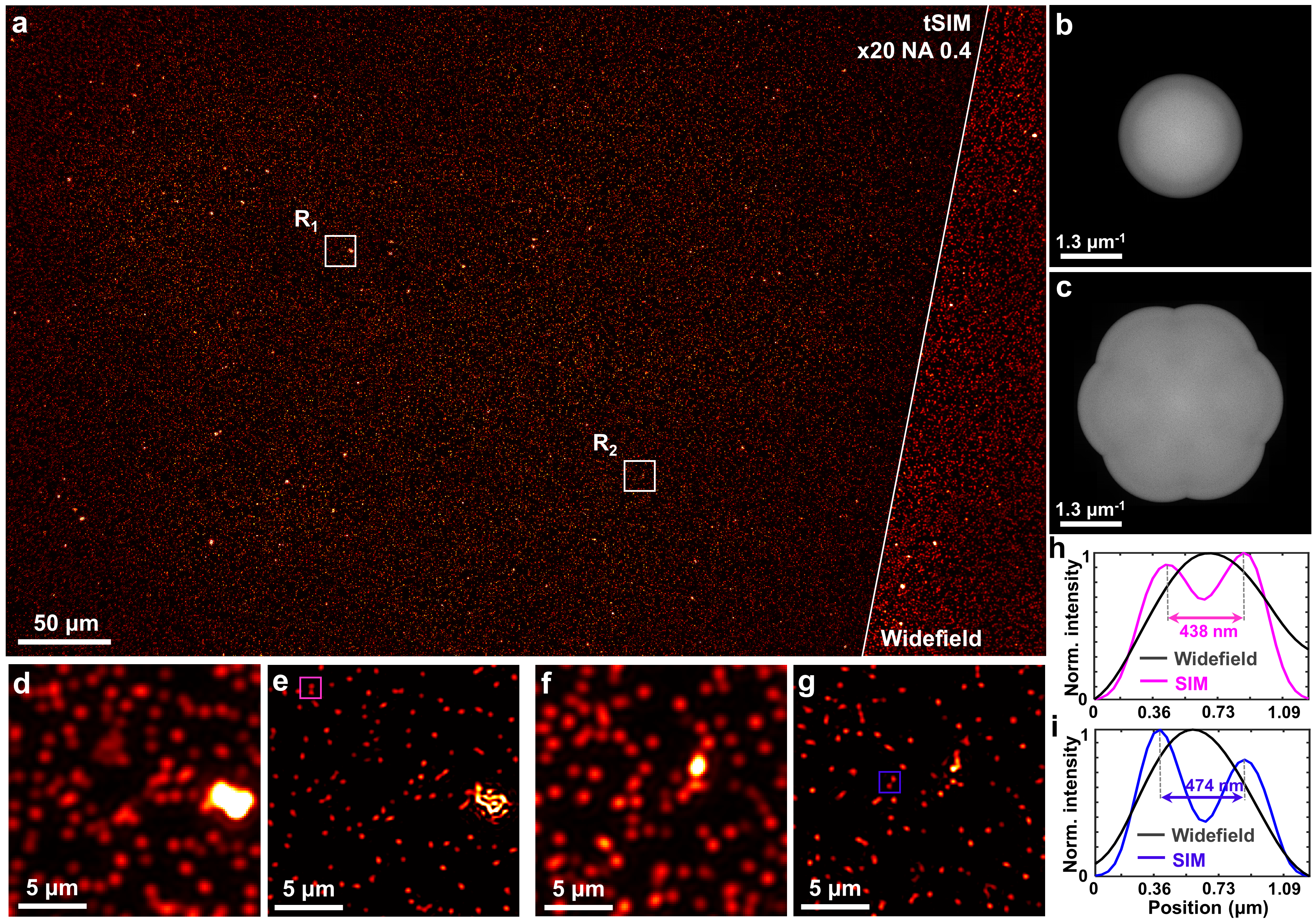}}
\caption{Experimental results of fluorescent beads: (a) single FOV tSIM reconstruction and widefield image of 200 nm fluorescent beads; (b) diffraction limited spectra; (c) reconstructed tSIM spectra; (d,f) magnified widefield images of white box in R$_1$ and R$_2$ regions, respectively; (e,g) closeup of the tSIM reconstruction in R$_1$ and R$_2$ regions, respectively; (h) intensity line-profiles of (d,e); (i) intensity line-profiles of (f,g).}
\label{fig:beadssim}
\end{figure}
A standard upright fluorescence microscope is used as the detection system which consists of a low-NA objective lens (PLN 20X/0.4, Olympus), matched tube-lens (U-TLU, Olympus), fluorescence filter (550 nm long-pass, Semrock) and a digital camera with rectangular sensor (Hamamatsu ORCA spark, 1920$\times$1200 pixels). The sparsely distributed fluorescent nano-particles of 200 nm diameter (Bangslabs) are used as the fluorescent sample sandwiched between two cover-glasses. The excitation and emission peak wavelength of the fluorophore in the beads is 540 nm and 600 nm, respectively. Employing low-NA collection enables imaging with the large FOV of 0.56 mm width and 0.35 mm height (captured by rectangular sensor). Following the conventional SIM approach, three raw moir\'e frames are acquired by translating the physical grating with a step of 0.55$\mu$m which leads to the phase-shifting of 0, 2$\pi$/3 and 4$\pi$/3 in the illumination pattern. The raw data for three different orientations are sequentially captured using three pair of tilt-mirrors in the hexagonal assembly. In total 9 frames (3 phases $\times$ 3 orientations) are computationally processed to achieve the final image. The open-source reconstruction algorithm `OpenSIM'\cite{lal2016structured} is applied to process the raw tSIM data. The reconstructed image of the fluorescent beads and corresponding diffraction limited widefield image are shown in the Figure \ref{fig:beadssim}(a). The power spectrum of the widefield image is shown in the Figure \ref{fig:beadssim}(b), whereas the collected tSIM spectra are shown in Figure \ref{fig:beadssim}(c). The details of the separated spectral lobes are shown in the Supplementary 4. The magnified views of the white-box region R$_1$ are shown in Figures \ref{fig:beadssim}(d,e), which correspond to the widefield and reconstructed tSIM image, respectively. The intensity line profile across the beads in the region marked by the magenta-box is shown in Figure \ref{fig:beadssim}(h). The zoom-in details of another region R$_2$ is displayed by the Figures \ref{fig:beadssim}(f,g) corresponding to the widefield and reconstructed image respectively. Figure \ref{fig:beadssim}(i) represents the line-scan profile across two fluorescent beads within the blue-box region. As the FOV is quite large, the analysis of several such zoom-in regions are shown in the Supplementary 3 to verify the homogeneous resolution enhancement over the entire FOV. Resolution enhancement is prominently visible from the intensity line profiles, where the continuous intensity distribution across two beads in the widefield image appears as two isolated particles after the tSIM reconstruction(\ref{fig:beadssim}(h,i)). The theoretical limit of the spatial resolution of the microscopy system is estimated to be 750 nm. It is thus confirmed that the separations above 430 nm are resolved by tSIM under conventional SIM conditions at a large FOV of 0.56mm$\times$0.35mm (restricted by the imaging sensor size).

\begin{figure}[t!]
\centering
{\includegraphics[width=1\linewidth]{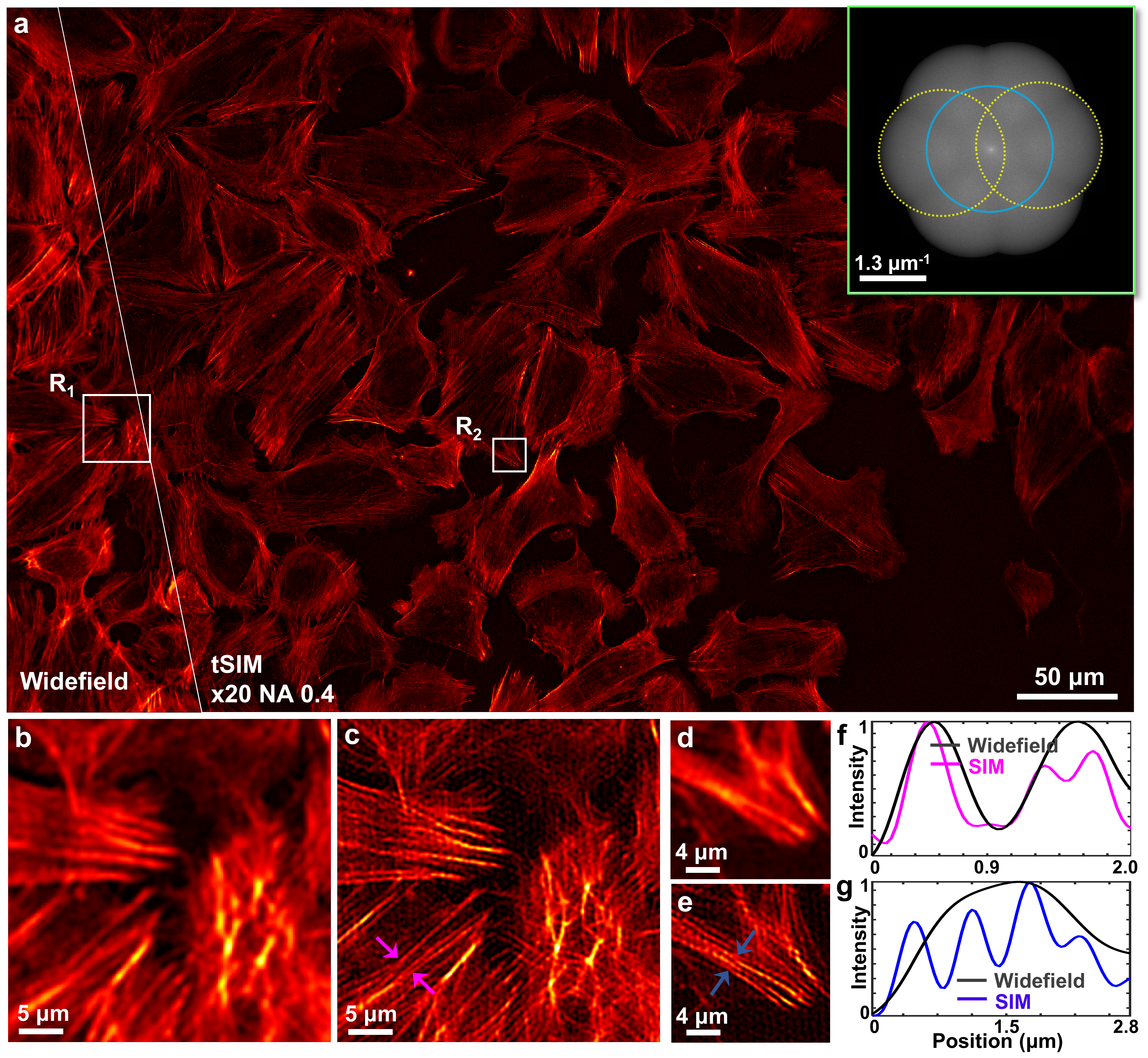}}
\caption{tSIM imaging of actin filaments in U2OS cells: (a) single FOV tSIM reconstruction vs. diffraction limited image with the excitation/emission wavelengths of 532 nm/560 nm and inset shows spectra of the reconstructed image; (b,d) diffraction limited images of regions R$_1$ and R$_2$; (c,e) reconstructed tSIM image of regions R$_1$ and R$_2$ respectively; (f,g) intensity profile of a line marked between the arrowheads in (c.e).}
\label{fig:actinsim}
\end{figure}
Further, we explore the feasibility of the technique for a real application through a biological specimen. U2OS (human bone osteosarcoma) cells are seeded  on glass coverslips, permeabilized, stained for actin filaments, and sealed with another glass coverslip on top. The 20x/0.4 objective lens is again employed here to simultaneously record more than 50 cells within the large FOV (0.56 mm width and 0.35 mm height). The sample imaging is performed in a region of interest where the concentration of the cells are moderate. The raw moir\'e frames are acquired under the identical imaging conditions as described for the case of the fluorescent beads. The reconstruction algorithm `OpenSIM'\cite{lal2016structured} is implemented in this example as well, with the widefield and corresponding reconstructed tSIM image shown in Figure \ref{fig:actinsim}(a). The inset of Figure \ref{fig:actinsim}(a) represents the spatial frequency spectra of the reconstructed tSIM image. The figures \ref{fig:actinsim}(b,d) represent the details from the white-box regions R$_1$ and R$_2$, respectively. The single FOV tSIM images corresponding to these regions are shown in Figures \ref{fig:actinsim}(c,e). The intensity line profile of Figure \ref{fig:actinsim}(c) (marked by the magenta arrowheads) is plotted in Figure \ref{fig:actinsim}(f). Similarly, another line profile of Figure \ref{fig:actinsim}(e) (marked by the blue arrowheads) is shown in Figure \ref{fig:actinsim}(g). The intensity line profile measurements of a few more regions are presented in the Supplementary 5. The features which appear as uniform distribution in the widefield image are resolved in the reconstructed image, where the spikes in the intensity profiles demonstrate the resolution enhancement of tSIM for the bio-samples.

\subsection*{tSIM imaging with high frequency illumination}
\begin{figure}[t]
\centering
{\includegraphics[width=1\linewidth]{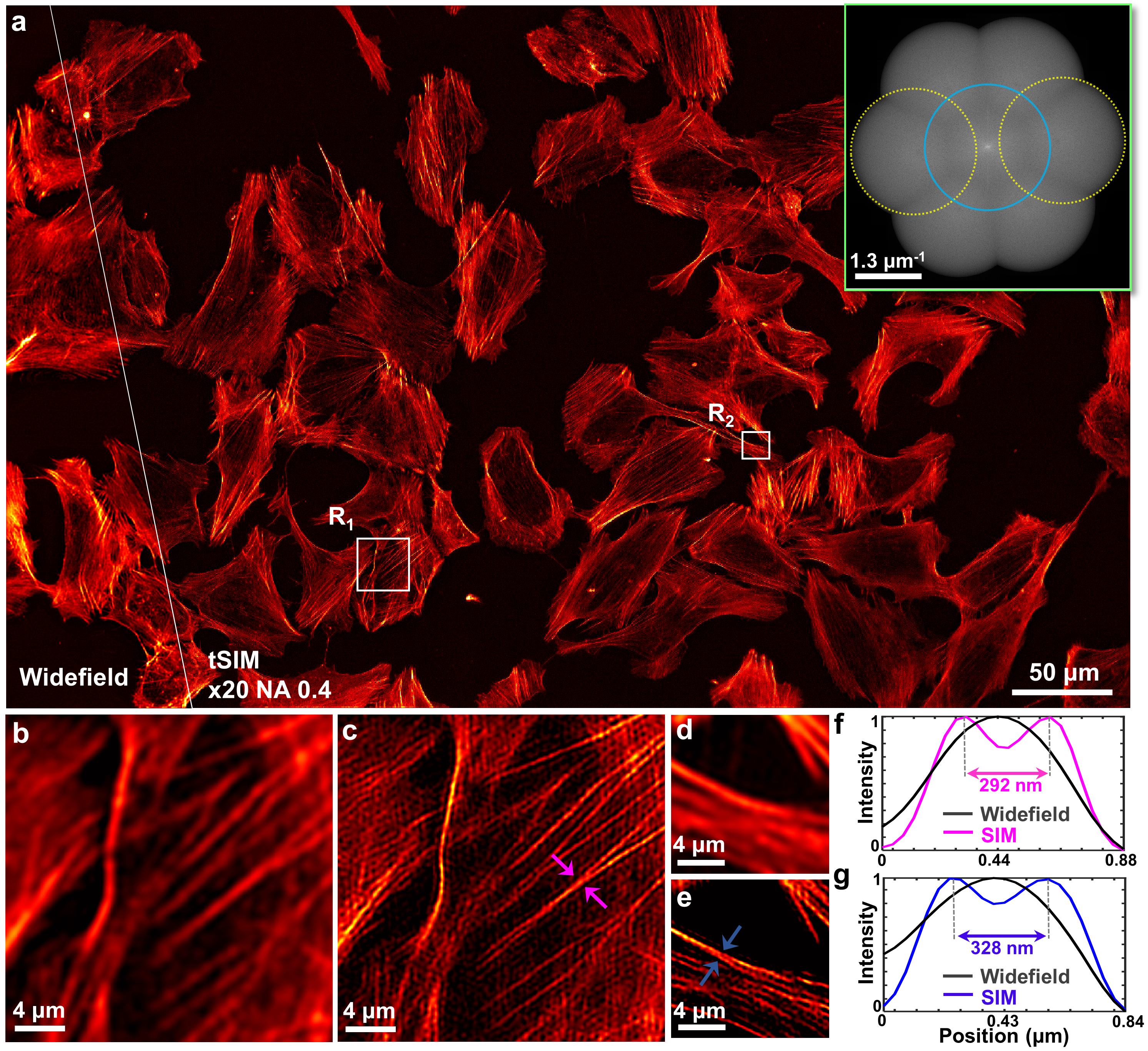}}
\caption{High-frequency tSIM imaging of actin-stained U2OS cell: (a) single FOV tSIM reconstruction and diffraction limited image with the excitation/emission wavelengths of 532 nm/560 nm and inset shows spectra of the reconstructed image; (b,d) widefield image of R$_1$ and R$_2$ regions respectively; (c,e) reconstructed image of R$_1$ and R$_2$ regions respectively; (f,g) intensity line profiles between the magenta and blue arrowheads.}
\label{fig:highfreq}
\end{figure}
So far we have shown tSIM results with traditional illumination conditions. Now, we present an example where tSIM surpasses the conventional linear SIM resolution gain of 2-fold, performed by implementing high frequency illumination with the pattern frequency outside the detection passband limit. The angle of the tilt-mirrors are tuned in such a way that the semi-angle of interference becomes $\theta$=35$^{\circ}$. This facilitates the generation of an interference pattern of 464 nm periodicity in the sample plane,  which is below the diffraction limit of the detection system and thus fringes are not possible to observe. 

The raw moir\'e data is acquired by keeping the same detection objective lens (20X/0.4) as well as the previously mentioned imaging parameters. As before, 9 raw frames (3 phases $\times$ 3 orientations) were recorded in total. However, the conventional reconstruction approaches \cite{lal2016structured,muller2016open} suffer to process the raw data in this particular case as the illumination frequency exists outside the passband support. This is overcome by utilizing a TIRF-SIM like reconstruction approach \cite{lal2016structured,samanta2021image}. The reconstructed and corresponding widefield image is shown in Figure \ref{fig:highfreq}(a). The inset of Figure \ref{fig:highfreq}(a) represents the frequency spectra of the reconstructed tSIM image. The off-center spectral components (marked with dotted yellow) are not overlapping which demonstrates the coverage to the broader spectral range due to high frequency illumination patterns. Figures \ref{fig:highfreq}(b,d) represents widefield image of two zoom-in regions R$_1$ and R$_2$ respectively. Their corresponding super-resolved images are shown by Figures \ref{fig:highfreq}(c,e). The intensity line profiles between the magenta arrowheads in Figures \ref{fig:highfreq}(b,c) and the blue arrowheads in Figures \ref{fig:highfreq}(b,c) are plotted in the Figure \ref{fig:highfreq}(f) and Figure \ref{fig:highfreq}(g) respectively. This leads to a resolution of $\sim$290 nm with the FOV of size 0.56mm$\times$0.35mm. Resolution enhancement of tSIM under different illumination configurations in terms of spectral support is presented in the Supplementary 6.

\subsection*{Scalable tSIM imaging}
All the previous discussion revolves around low-NA collection, large FOV imaging, tunable illumination patterns and corresponding super-resolution. However, it is mentioned that the overall resolution appears from the joint contributions of illumination and detection. Finally, the super-resolution capability of the tSIM utilizing the high-magnification/high-NA detection objective is demonstrated. We switch to 100X/1.3 oil immersion (Olympus) objective lens for the collection of the fluorescent signal while keeping the other imaging components (i.e. fluorescent filters, camera sensor) same as described for the previous cases. The same sample (U2OS cells stained with alexa-fluor 532) is used here also. 
\begin{figure}[t!]
\centering
{\includegraphics[width=0.8\linewidth]{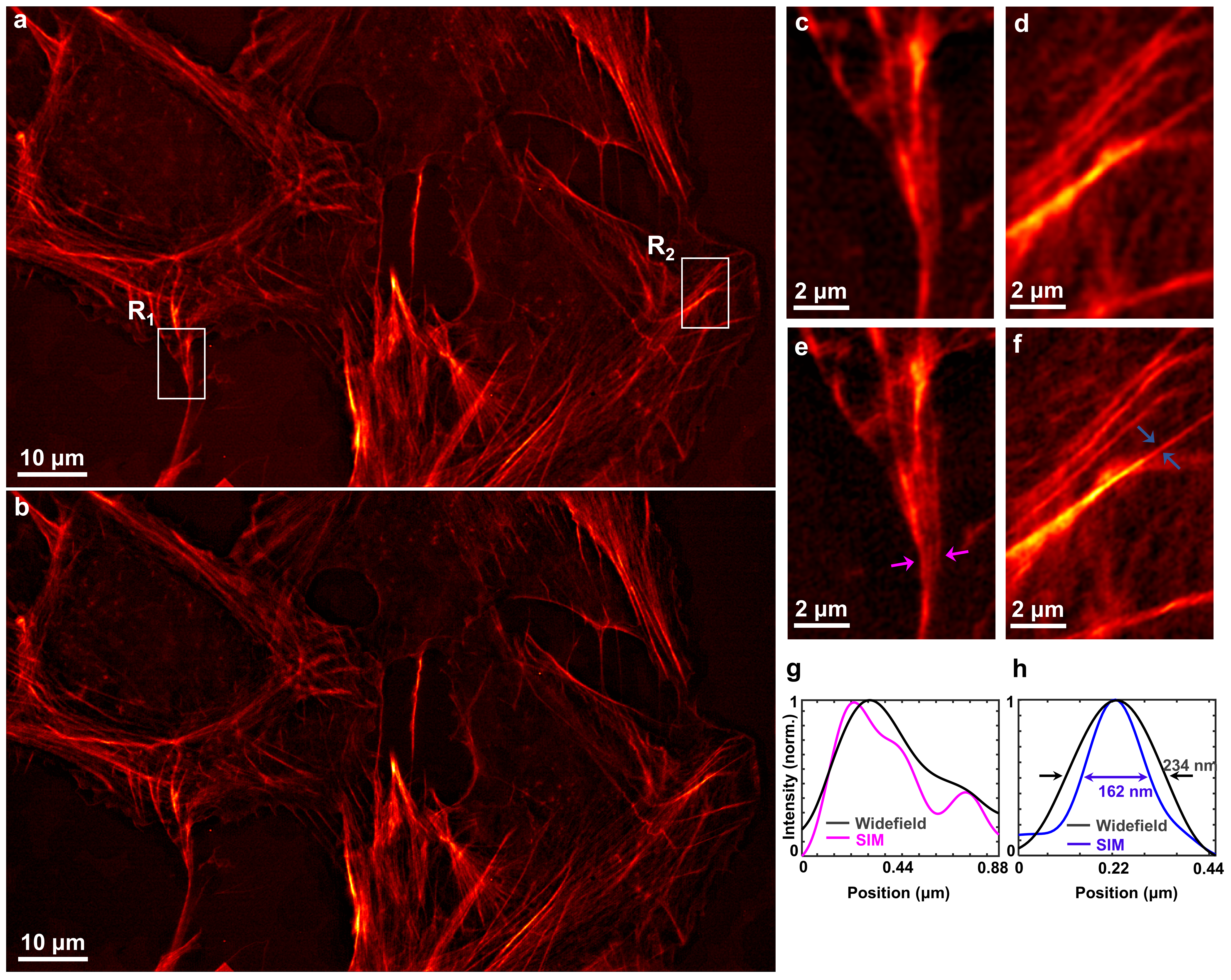}}
\caption{Actin of U2OS cell stained with alexa 532: (a) widefield image; (b) SIM reconstruction image; (c,d) magnified widefield images of white box in R$_1$ and R$_2$ regions respectively; (e,f) closeup of the SIM images in R$_1$ and R$_2$ regions respectively; (g) intensity line-profiles of (c,e) shown by magenta colour; (h) intensity line-profiles of (d,f) shown by blue colour.}
\label{fig:highNA}
\end{figure}
The semi-angle of interference is $\theta$=$45^{\circ}$ that leads to generate interference fringes with spacing 376 nm in the sample plane. This leads to a illumination case similar to conventional SIM with high-NA and the raw data are recorded following the same way. The widefield image is shown in the Figure \ref{fig:highNA}(a) whereas corresponding tSIM image is depicted in the Figure \ref{fig:highfreq}(b). Figures \ref{fig:highfreq}(c,e) represent the magnified views of the widefield image in the regions R$_1$ and R$_2$ respectively. The close-up view of the tSIM images are shown in Figures \ref{fig:highfreq}(d,f) respectively. The intensity line profiles between the magenta arrowheads in Figures \ref{fig:highfreq}(c,e) and the blue arrowheads in Figures \ref{fig:highfreq}(d,f) are plotted in the Figure \ref{fig:highfreq}(g) and Figure \ref{fig:highfreq}(h) respectively. Thus, employing high-NA objective improves the resolution of the detection system and applying tSIM brings the resolution scale down to $\sim$160 nm. However, the super-resolution in this particular case appears at a cost of the FOV i.e. 0.12mm$\times$0.70mm. The super-resolution capability of the tSIM with various illumination and detection configuration have been shown in the Table \ref{tab:table}. 
\begin{table}[htbp!]
	\caption{Super-resolution achieved by tSIM under different illumination-detection conditions}
	\centering
	\begin{tabular}{lllllll}
		\toprule
		Excitation/ & Interference & Illumination     & Detection    & Field of view & Widefield   & tSIM \\
		emission (nm) & angle ($\theta$) & effective NA & objective M/NA & (mm$\times$mm) & resolution  & resolution \\
		\midrule
		532/600 & 19$^{\circ}$ & 0.32  & 20X/ 0.4   & 0.56$\times$0.35  & 750 nm  & $\sim$430 nm\\
		532/560 & 35$^{\circ}$ & 0.57  & 20X/ 0.4   & 0.56$\times$0.35 & 700 nm  & $\sim$290 nm\\
		532/560 & 45$^{\circ}$ & 0.71  & 100X/ 1.3  & 0.11$\times$0.07 & 215 nm  & $\sim$160 nm\\
		\bottomrule
	\end{tabular}
	\label{tab:table}
\end{table}

\section*{Conclusion and discussion}
In conclusion, we demonstrate the proof-of-concept experimental results of tilt-mirror assisted transmission structured illumination microscopy which makes the illumination and the detection optics flexible to configure them independent to each other. The tilt-mirror facilitates to generate illumination patterns with tunable frequencies, which are of potential importance in order to provide the resolution gain depending upon the requirement. Employing the low-NA detection, a single large FOV image of size 0.56mm$\times$0.35mm (primarily limited by the sensor boundary) is obtained and we first demonstrate $\sim$1.7-fold resolution improvement (conventional SIM) through tSIM over the diffraction limit. The experimental verification is performed for both the fluorescent beads as well as for biological samples. Further by exploiting the high frequency illumination pattern (frequency peak outside the detection passband), a $\sim$2.4-fold resolution enhancement is achieved via tSIM with the same FOV. It is observed that the traditional reconstruction fails to process the raw data for the cases when the illumination frequency lies outside the detection passband. We have implemented TIRF-SIM like reconstruction \cite{lal2016structured} approach for our reconstruction, however, there are possibilities to explore blind \cite{yeh2017structured,samanta2021blind} approaches which does not require the prior knowledge of the illumination patterns. Moreover, the deep-learning based \cite{jin2020deep,ling2020fast} might offer potential application in the tSIM reconstruction. The proof-of-principle experimental demonstration of tSIM for the fluorescent beads provides the perfect result. However, the reconstruction of the actin filaments contains some sort of honeycomb artifacts \cite{demmerle2017strategic} which is due to photo-bleaching and poor signal to noise ratio. Such reconstruction imaging artefacts are common in standard SIM image, however, the fluorescence is bright and stable for the beads sample and such artefacts were not observed. The feasibility of tSIM is justified with high-NA detection objective lens with the resolution down to $\sim$160 nm by compromising with the FOV. 

The system is very compact in comparison to the typical SIM systems. Employing high refractive index medium into the illumination configuration, sub-diffraction fringe patterns which cannot be realized in free-space optics. This might have potential application to further enhance the resolution of tSIM. Different excitation sources can be coupled into the system to probe different fluorophores for the multi-colour SIM imaging \cite{markwirth2019video,hinsdale2021high}. Nevertheless, there is a scope to extend the proposed technique suitable for the 3D-SIM \cite{schermelleh2008subdiffraction,shao2011super} imaging by allowing the central beam into the tSIM architecture. In conventional SIM systems, the illumination and the collection is coupled and thus to obtain the small fringe period (high-spatial frequency), the two interfering beams should hit the edge of the back aperture of a given objective lens. Thus, every time a new objective lens is used for imaging, the optical arrangement must be changed to utilize the highest spatial frequency supported by the different objective lenses. Consequently, the commercial SIM system usually comes with a pre-defined single objective lens of high magnification and N.A. This restricts the FOV of commercial SIM system to about $50\times50 {\mu}m^2$. The tSIM supports both the large and the user-defined FOV by simply using swapping to another detection objective lens without influencing the illumination light path. The large FOV support of tSIM will find applications where high-throughput is essential such as in the histopathology \cite{villegas2020visualizing}. Similarly, tSIM will benefit neuroimaging applications where data must be correlated both in space and time by imaging large areas \cite{opstad2020waveguide}. 

\section*{Appendices}
\subsection*{Preparation for fluorescent-beads sample}
The coverslip was rinsed with diluted acetone and cleaned with deionized water. Next, it was treated with isopropyl alcohol solution and again cleaned with deionized water. Then, it was dried with the compressed nitrogen gas. A small (0.5 $\mu l$) amount of 200-nm-diameter fluorescent bead solution (Suncoast yellow, FSSY002 Banglabs) was pipetted out and poured gently onto the cleaned coverslip. The coverslip was then kept at an inclined position and then allowed to dry completely. In this kind of drying, gravity pulls the droplet and the upper edge down during the liquid evaporation, which reduces the `coffee ring effect'. A tiny drop of glycerol was placed on another coverslip of bigger length; touching one edge of the coverslip containing beads with the bigger coverslip, the other edge was lowered slowly until it became completely flat on the glycerol. During this process, the sample side of the small coverslip faced towards glycerol so that the sample became sandwiched between two coverslips. The glycerol was allowed to spread out without pressing and the boundary between two coverslips was sealed with nail paint. After drying the paint, the sample was put under microscope for imaging. 

\subsection*{Cell culture and labeling}
Osteosarcoma U2OS cells were cultured in Dulbecco's modified eagle medium (DMEM, high glucose, Sigma-Aldrich) supplemented with 10\% fetal bovine serum (Sigma-Aldrich) and 1\% penicillin/streptomycin (Sigma-Aldrich) at 37 degree with 5\% CO2. Cells (15-20\% confluence) were seeded on \#1.5 glass coverslips( 22X22mm, VWR) and incubated for 1 day before imaging. Then cells were washed in PBS, fixed with 4\% formaldehyde (FA, ThermoFisher) in PBS for 10 minutes, washed three times in PBS, and followed with 0.1\% TritonX-100 (Sigma-Aldrich) in PBS for 10 minutes for permeabilization, and washed three times in PBS. Cells were then incubated with Alexa FluorTM 532 phalloidin (1:50 in PBS, ThermoFisher) for 20 minutes at room temperature (RT) to label f-actin in the cells. Three times wash in PBS before mounting with ProLong$^{\mathrm{TM}}$ Glass Antifade Mountant (ThermoFisher) on glass slide/coverslips (24x50mm, VWR) at RT overnight. Samples are then ready for the imaging.

\subsection*{Experimental setup and data acquisition}
The raw moir\'e frames were recorded using an upright modular microscope (BXFM, Olympus). The emission filter used for the imaging is the 550 nm long-pass fluorescent filter (Semrock). The rotation of grating was done using a motorized rotation stage (Holmarc). The physical grating was mounted on a single-axis translation stage and this stage including grating were attached together with the rotation stage. The linear piezo-translation stage (NF15AP25/M, Thorlabs) is operated along a particular direction to introduce suitable phase-shifting to the illumination pattern and a sequence of frames are recorded correspondingly. The acquisition process is repeated for three different orientations by revolving the grating around the optic axis using the rotation stage (MPR-SSHR-50, Holmarc). 

\subsection*{Image reconstruction}
The computational image processing i.e. the reconstruction steps are performed only using MATLAB R2020b executed in a computer workstation running Windows 10 (Intel Xeon(R) W-2175 CPU, 2.50 GHz, RAM 64GB). First, the raw dataset is drift corrected through a custom-written algorithm. Then open-source package (Open-SIM) is employed to obtain the super-resolution image from the drift-corrected data. The associated code called TIRF-SIM in the same package is utilized for the reconstruction of data where the illumination frequency lies beyond the detection passband. 

\section*{Acknowledgements}
The authors would like to thank Hong Mao, Deanna Wolfson and Florian Str\"{o}hl for assistance with the sample preparation and discussions about the SIM reconstruction. This work is supported by the Indo Norwegian Collaborative Program (Project Number - INCP 2014/10024), the Research Council of Norway, INTPART project (No: 309802), University Grants Commission (UGC, India), Defence Research and Development Organisation (RP03707G, DRDO).

\section*{Author contributions}
J.J. and B.S.A. conceived the idea. K.S., J.T., and A.A. have designed the tilt-mirror mount, prepared the samples, performed the experiments as well as reconstructions. J.J. and B.S.A. supervised the whole work. All the authors participated in the discussions to shape the research, analysis, and to construct the manuscript.

\section*{Conflict of interest}
A patent has been filed based on the tSIM technique.

\bibliographystyle{unsrt}
\bibliography{references}  




\clearpage
\section*{Supplementary information}
\subsection*{Supplementary 1: Tilt-mirror angle-tuning and volumetric interference}
The physical grating employed for diffraction contains 600 lines per mm i.e. the periodicity is 1.666-micron. The central order is blocked and the first diffraction orders (+1, -1 orders of equal intensity) are allowed to interfere and to generate the interference pattern at the sample plane. The first order diffracted beams come at an angle of ($\alpha$) is 18.6$^{\circ}$ with respect to the optics axis ($z$) and the semi-angle of interference $\theta$ = $2\delta+\alpha$; where $\delta$ is the mirror tilt-angle with respect to the optic axis. The maximum possible $\delta$ is 30$^{\circ}$ with the current design, it can also be customized depending upon the requirements.
\begin{figure}[htbp!]
\centering
{\includegraphics[width=0.8\linewidth]{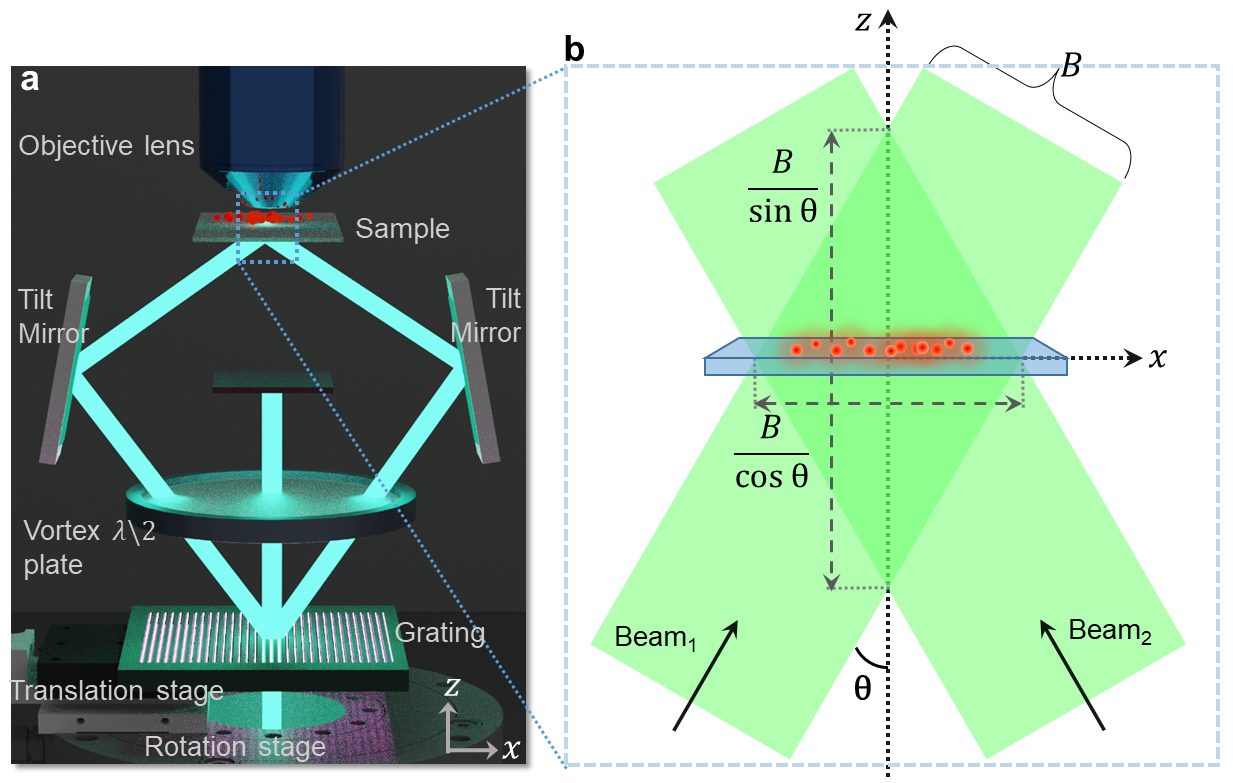}}
\caption{(a) Side-view of the graphical experimental setup; (b) volumetric interference and span of illumination region.}
\label{fig:S1}
\end{figure}
So, the pattern frequency can be modified by changing the diffraction angle (i.e. periodicity) of the physical grating or the tilt-angle of mirror. For a standard fixed grating, the tunability of the illumination appears only by tuning the mirror tilt-angle. The volumetric interference pattern is generated in region with $\frac{B}{{\cos \theta}}$ lateral and $\frac{B}{{\sin \theta}}$ axial spread. The span of the region depends on the beam diameter ($B$) and semi interference angle ($\theta$).

\clearpage
\subsection*{Supplementary 2: Nitty-gritty details of the experimental setup}
The mirror assembly consists of six mirror facets arranged symmetrically around the central (z) axis. The mirror-mount is made of metallic aluminium (Al) and hence provide compactness and robustness to the illumination pattern. Each mirror in the assembly is inclined at an angle $60^{\circ}$ w.r.t. horizontal direction.
\begin{figure}[htbp!]
\centering
{\includegraphics[width=1\linewidth]{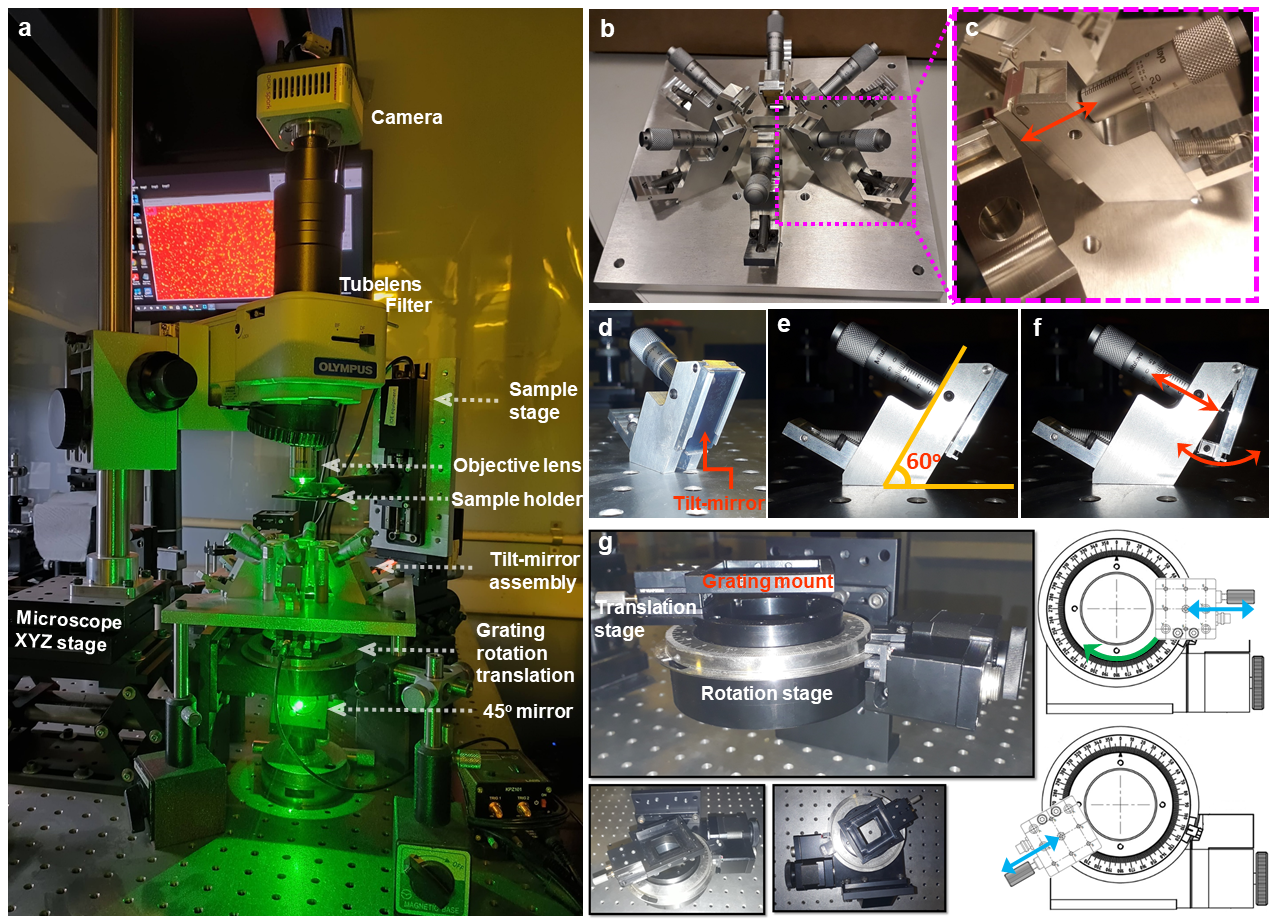}}
\caption{(a) Photograph of the real experimental setup; (b) metallic hexa mirror arrangement with the micrometer assisted angle tuning; (c) magnified view of a single tilt-mirror unit; (d) perspective view of single mirror unit; (e-f) angle tuning scheme of a single mirror facet; (g) coupled translation and rotation stage for the change of orientation and phase-shifting of the physical grating.}
\label{fig:S2}
\end{figure}
A micrometer screw-head is attached to each of the mirror facet, the linear movement of the screw head tunes and determines the tilt angle of that mirror. The whole system is built up vertically and imaging is done using a modular upright microscope kept on a $xyz$ stage. The physical grating below the hexa-mirror assembly is mounted on single axis piezo translation stage which is further coupled with a motorized rotation stage. The combined opto-mechanical movements of the physical grating by the linear and rotation stages provide the desired phase-shifting and change in the orientation of the illumination patterns. The central beam blocker and vortex half wave plate are mounted in a 3D printed holder and placed between the physical grating and multi-mirror assembly. The sample mount is attached to a $xyz$ stage to change the sample position according to the volumetric interference pattern.

\clearpage
\subsection*{Supplementary 3: Analysis of the SIM result for fluorescent beads}
The tSIM imaging of the fluorescent beads under conventional case is presented earlier. The homogeneity in the large FOV imaging with better resolution is confirmed here with the magnified views of six different regions.
\begin{figure}[htbp!]
\centering
{\includegraphics[width=1\linewidth]{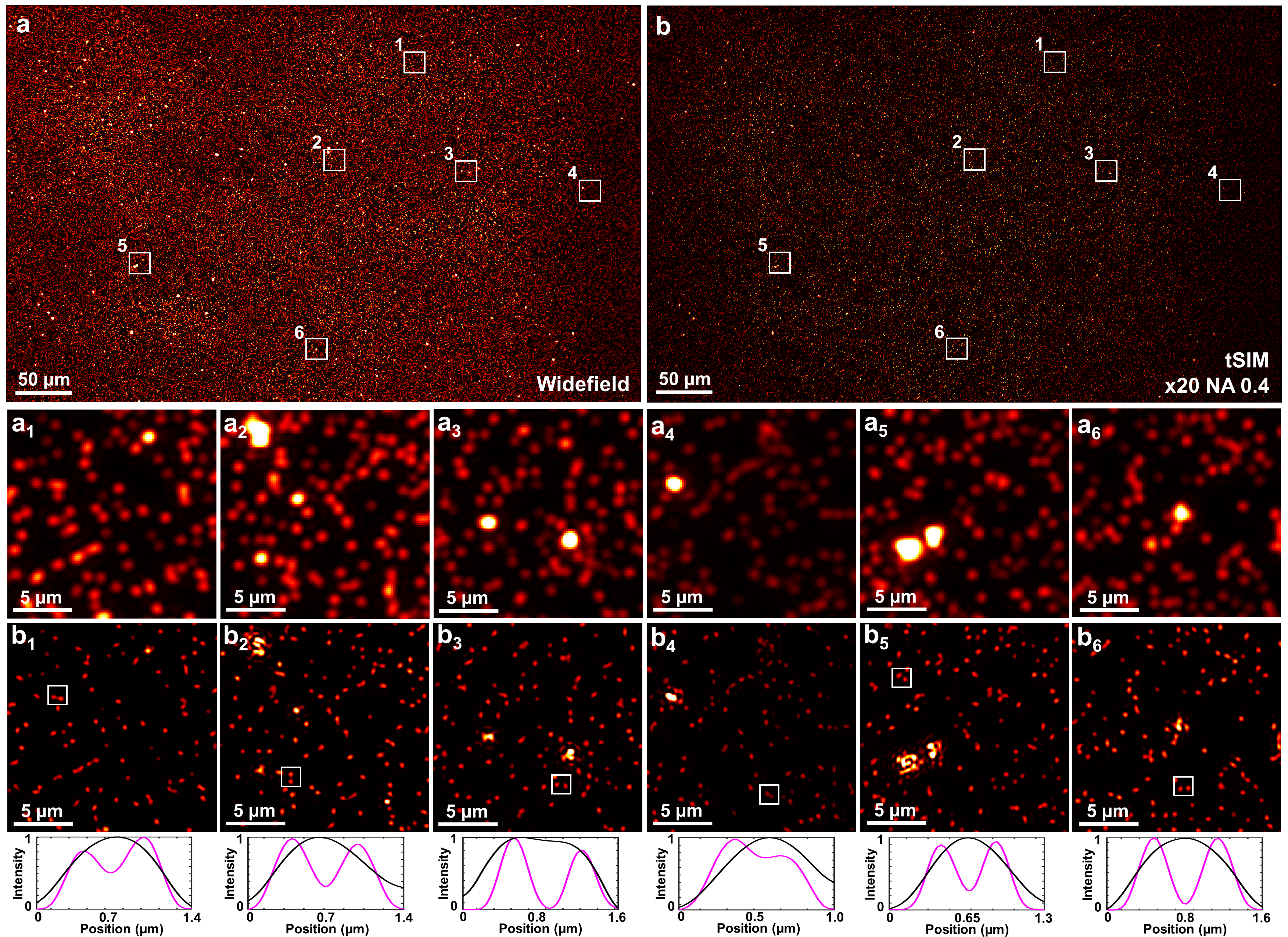}}
\caption{Experimental results of fluorescent beads: (a) diffraction limited image; (b) SIM reconstructed image; (a$_1$-a$_6$) magnified views of (a) in different regions; (b$_1$-b$_6$) magnified views of (b) in different regions.}
\label{fig:S3}
\end{figure}
The intensity line profiles across the beads in the selected box are presented by the magenta and black coloured plots in the lower panel. This proves the resolution enhancement through tSIM in each of the zoom-in regions.

\clearpage
\subsection*{Supplementary 4: Power spectral support for the beads data}
The power spectra for the fluorescent beads data are shown here. The upper panel (figures a, b, c) show the off-center frequency components for three different orientations and the peak frequency is marked by the cyan-coloured circles. 
\begin{figure}[htbp!]
\centering
{\includegraphics[width=0.8\linewidth]{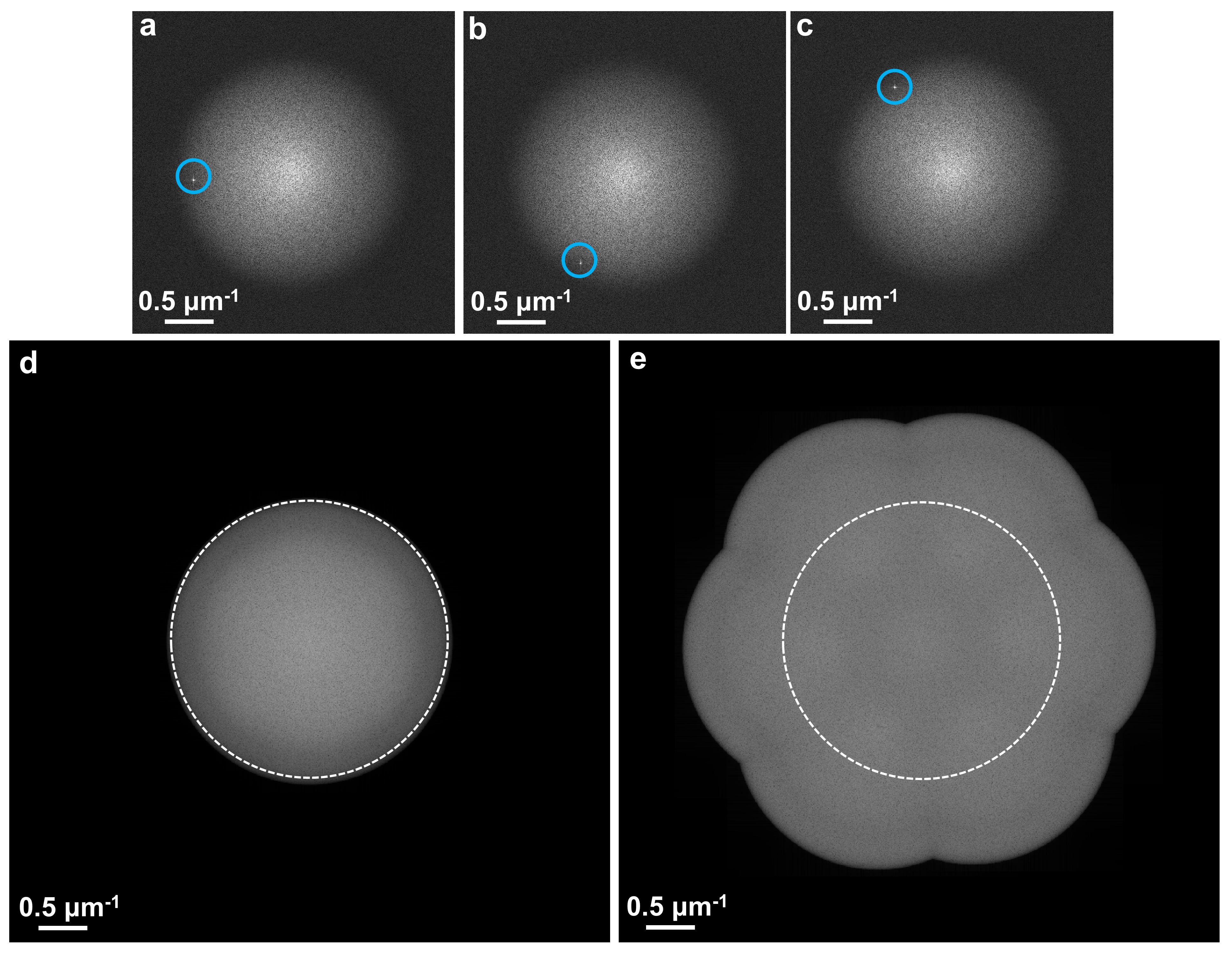}}
\caption{2D spatial frequency spectra of SIM data for fluorescent beads: (a-c) three separated spectral bands used for the reconstruction process where peaks corresponding to the illumination frequency and orientation are marked with the circles; (d) diffraction limited spectra; (e) reconstructed SIM spectra.}
\label{fig:S4}
\end{figure}
The lower panel (figures d, e) represent the diffraction limited spectra as well as tSIM spectra respectively for the conventional illumination case ($\theta = 19^{\circ}$). The white circles in both the cases correspond to the power spectral support of the diffraction-limited image.

\clearpage
\subsection*{Supplementary 5: Analysis of the SIM result for the actin}
The tSIM imaging of actin filaments of U2OS cell under conventional SIM illumination scheme ($\theta = 19^{\circ}$) is previously presented and analyzed with some magnified regions. 
\begin{figure}[htbp!]
\centering
{\includegraphics[width=1\linewidth]{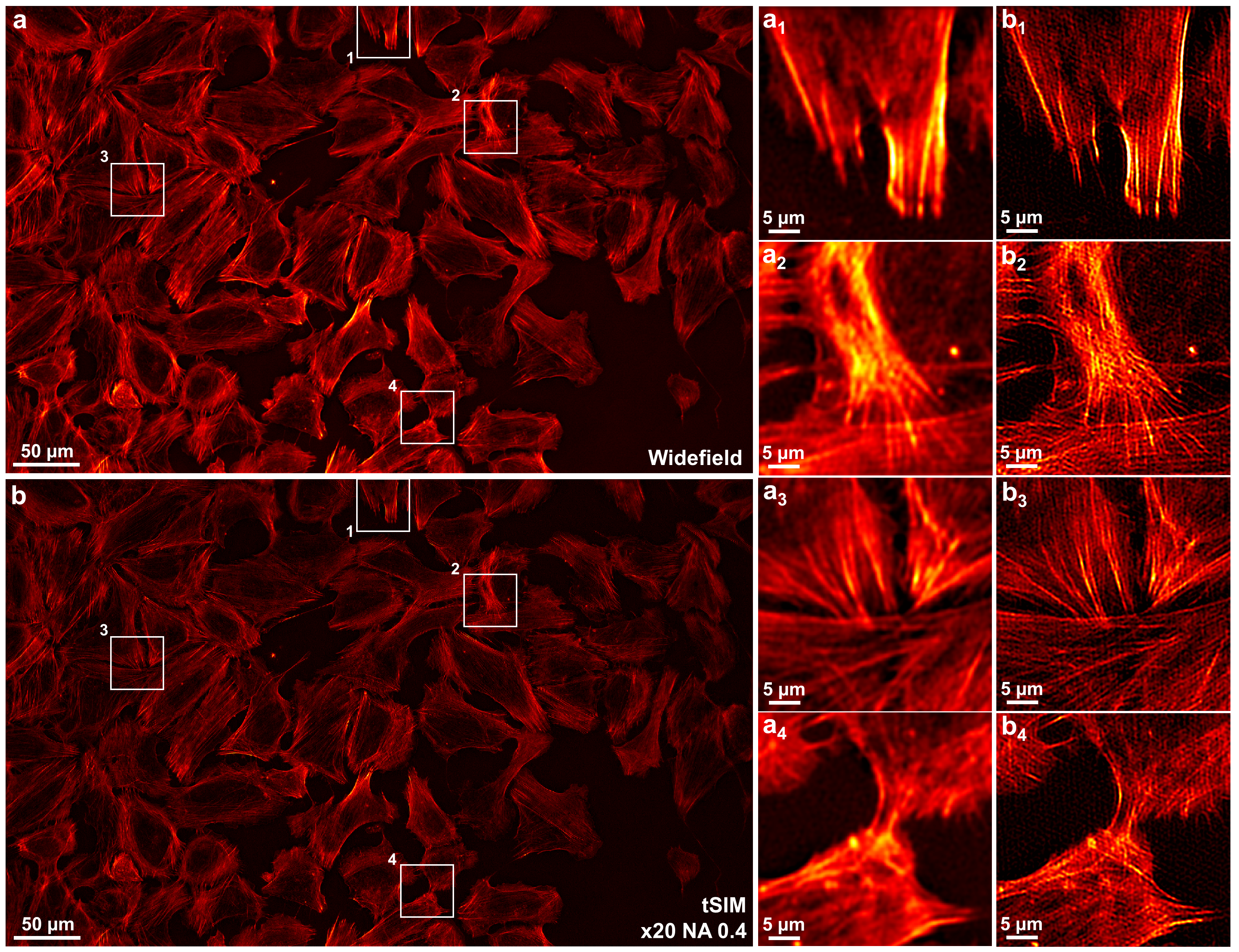}}
\caption{Experimental results of actin filaments of U2OS cell stained with alexa-fluor 532 fluorescent tagging: (a) diffraction limited image; (b) SIM reconstructed image; (a$_1$-a$_4$) magnified views of (a) in different regions; (b$_1$-b$_4$) magnified views of (b) in different regions.}
\label{fig:S5}
\end{figure}
Here, the homogeneity in the large FOV imaging with better resolution is verified with the zoom-in views of a few more selected regions of the widefield as well as tSIM image, which visually confirms the actin filament in the reconstructed tSIM image to be nicely resolved.

\clearpage
\subsection*{Supplementary 6: Spectral support of tSIM under different illumination configurations}
The spatial frequency support of the tSIM results for different illumination configurations using 20X/0.4 detection objective are shown here. The center circle (solid cyan colour) represents diffraction limited support of the detection system. 
\begin{figure}[htbp!]
\centering
{\includegraphics[width=1\linewidth]{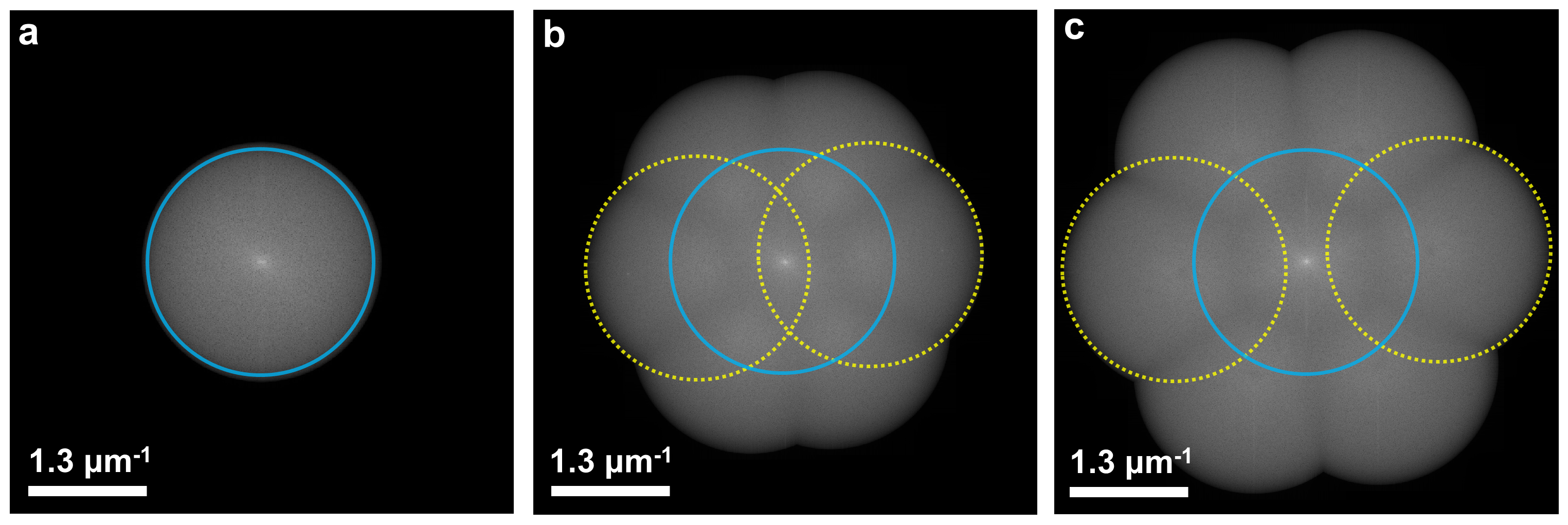}}
\caption{Spatial frequency support: (a) diffraction limited image; (b) tSIM reconstructed image for conventional illumination; (c) tSIM reconstructed image for high frequency illumination.}
\label{fig:S6}
\end{figure}
The off-center circles (dotted yellow colour) corresponds to the high frequency spectral components. The tSIM spectra for two different illumination scenarios ($\theta = 19^{\circ}$ \& $\theta = 35^{\circ}$) are shown for the comparative demonstration of passband extension due to frequency tuning. In first case ($\theta = 19^{\circ}$), the illumination pattern frequency lies within the detection passband and the other case ($\theta = 35^{\circ}$) deals with illumination pattern with frequency outside the detection passband.

\end{document}